\documentclass{INTERSPEECH2023}

\interspeechcameraready 

\let\underscore=\_
\def\_{\checkmath_\underscore}
\def\checkmath#1#2{\ifmmode\def\next##1{#1{\rm##1}}\else\let\next=#2\fi\next}
\let\circonflex=\^
\def\^{\checkmath^\circonflex}

\def\fpr{\ensuremath{p\_{fp}}}
\def\fnr{\ensuremath{p\_{fn}}}
\def\?{\hphantom{0}}

\title{Speaker and Language Change Detection using Wav2vec2 and Whisper}
\name{Tijn Berns$^1$, Nik Vaessen$^1$, David A. van Leeuwen$^{1,2}$}
\address{
  $^1$Radboud University Nijmegen, The Netherlands\\
  $^2$SpraakLab, The Netherlands
}
\email{tijnberns.berns@student.ru.nl, nik.vaessen@ru.nl, david.vanleeuwen@ru.nl}

\begin{document}

\maketitle

\begin{abstract}
    We investigate recent transformer networks pre-trained for automatic speech recognition for their ability to detect speaker and language changes in speech.  We do this by simply adding speaker (change) or language targets to the labels.  For Wav2vec2 pre-trained networks, we also investigate if the representation for the speaker change symbol can be conditioned to capture speaker identity characteristics.  Using a number of constructed data sets we show that these capabilities are definitely there, with speaker recognition equal error rates of the order of 10\,\% and language detection error rates of a few percent.  We will publish the code for reproducibility. 
\end{abstract}

\section{Introduction}

Speaker Change Detection (SCD) is the task of finding the moments in a conversation where one speaker stops talking and another starts.  For some applications, such as speaker diarization, this task can act as a mere initialization of finding speaker clusters that are later refined---but it can also be a side effect of a diarization algorithm.  For other applications, such as online low latency generation of close captioning titles for the hearing impaired, it may add to the ease of understanding the conversation if speaker turns are indicated visually, e.g., by color of the captions.  This paper is about speaker change detection per se, and we restrict ourselves to the case of non-overlapping speech.  

The moments of speaker change are not necessarily always well-defined: in order of increasing overlap between speakers, we may have~\cite{Heldner:2010} \emph{gaps}, with a measurable silence between the speaker turns and \emph{speech overlap}, when the next speaker starts speaking before the previous is finished its turn.  Further, there are \emph{pauses} within a speaker turn and overlapping \emph{backchannels} from another speaker such as ``hmm'', ``uh-huh'' and ``yes'', where there isn't really a change of speaker.  
For a readable automatic transcription, it is important to get the words (including those near the speaker changes) and the speaker changes right, while representing the amount of overlap is of secondary importance. 
We therefore investigate in this paper how well a state-of-the-art automatic speech recognition (ASR) system can be extended with the task of finding speaker changes.  Both linguistic and acoustic features can be of use to finding speaker changes, and modern transformer-based ASR architectures can model both knowledge sources. 

Traditionally SCD algorithms were based purely on the acoustics, in~\cite{Chen:1998} the Baysian Information Criterion was introduced to assess a potential speaker change in terms of the benefit in the log likelihood if two separate models were used before and after the split, vs a single model for the combined segment.  A penalty was needed to compensate for the difference in the number of parameters, but tuning the weight of this penalty was considered a weakness, that~\cite{Ajmera:2004} cleverly circumvented by fixing the number of model parameters when going from a single to two models.  In the neural era, \cite{Yin:2017} applied an LSTM for the sole task of SCD, labelling individual frames with a speaker change boolean, after convolving the single speaker change labels with a unit block function to account for class imbalance.  In~\cite{Xia:2022}, a transformer transducer was trained to produce speaker changes as an output symbol in the target transcription, combining ASR and SCD tasks in a single model.  They further combine this with a speaker encoder and spectral clustering to carry out speaker diarization.  

Our contributions are the following.  We build upon the work of \cite{Yin:2017, Xia:2022}, but simply finetune existing ASR networks, Wav2vec2~\cite{Baevski:2020} and Whisper~\cite{Radford:2022}.  We further investigate if we can learn a speaker change symbol \emph{SC} for which the corresponding network representation functions as a speaker embedding for the following segment, so that speaker recognition can be carried out.  Finally, we show that speaker change labels can be combined with language identification labels in a multilingual setup. 

\section{Speaker Change Detection approach}

Our basic approach to speaker change detection is to finetune a network (pre-)trained for an ASR task where we include speaker (change) information as special labels.  Our default workhorse will be Wav2vec2~\cite{Baevski:2020}, for which pretrained models and code are readily available, and for which it has been shown that it can be finetuned to carry out various other tasks than ASR, such as emotion~\cite{Pepino:2021}, language~\cite{Tjandra:2021} and speaker recognition~\cite{Vaessen:2022}.  We carry out several experiments, demanding progressively more from a special target label that we will denote as \emph{SC}, for \emph{speaker change}.  For Wav2vec2, we extend the standard set of 28 letter targets, a--z, apostrophe and word-end (space), with SC.  This symbol is provided in the transcriptions, but does not have an explicit alignment---this will be implicitly found by the CTC training~\cite{Graves:2006}.  We trust that the expressiveness of the encoding transformer~\cite{Vaswani:2017} can deal with the fact that there is no ``single speech frame'' associated with the speaker change, but rather two sequences of frames that are simultaneously aligned with other letter targets.  We will have the following training conditions for SC:
\begin{description}
  \item[Speaker Separator] Here SC is placed between the transcription of the previous and the following speaker
  \item[Speaker Announce] Here SC is placed at the beginning of a segment of speech uttered by a single speaker.  This is the same as `speaker separator', except that an additional SC is placed at the beginning of each concatenated utterance. 
  \item[Speaker Label] This is the same as `Speaker Announce', but different target symbols are used for different speaker identities, so instead of a single target `SC' we have multiple targets `S1', `S2', \dots, the set of which functions as SC. 
  \item[No Speaker Changes] Contrasting the above conditions, all SC labels are removed, but the acoustic data remains the same.
  \item[Language Label] In the multilingual training condition, we use a language label (NL, DE, FR) to indicate both a speaker change and the language spoken in the following segment. 
\end{description}

The \emph{speaker separator} and \emph{speaker announce} are very similar, the former having a streaming application in mind where the latter is a precursor to the \emph{speaker label} and \emph{language label} experiments.  With \emph{speaker label} the idea is that the classification head functions more or less similar to the output head in contemporary speaker recognition systems~\cite{Snyder:2018, Desplanques:2020}, except that the loss function is cross entropy (through CTC) and competing with letter targets, rather than a more advanced Additive Angular Margin loss~\cite{Deng:2019}.  In the \emph{Language Label} condition, we have three extra targets for Wav2vec2, NL, DE and FR, instead of only SC. 

We also carry out experiments with Whisper~\cite{Radford:2022}, a more recent transformer encoder-decoder architecture that comes with pre-trained multi-lingual ASR models.  Here, we expect that the expressiveness of the decoder network~\cite{Chan:2016} can carry out the segmentation tasks, so
we simply use SC/NL/DE/FR labels as separate target words, without changing the tokenizer.  

\subsection{Data sources}

For this approach to work, we need to train or finetune with ASR transcriptions, as the transcriptions help in aligning the SC labels to actual times.  This puts a rather high demand on the data to use for training, as we need both transcriptions and speaker tags, preferably in a multiparty setting.  Because of the exploratory nature of this paper, we settle for constructing several datasets using standard datasets used in ASR training, similar to~\cite{Heo:2022}, by concatenating segments from the source datasets.  We use the following datasets for training various models:
\begin{description}
  \item[CGN-multispeaker] From the Dutch CGN~\cite{Oostdijk:2003}, we use the Northern and Southern Dutch variants, broadband components \emph{f}, \emph{i--l}.  This also was the training material for the NBest-2008 evaluation for Dutch ASR~\cite{N-best:2009}. 
  \item[LS-multispeaker] For English SC and speaker recognition we use the Librispeech clean 100 hours subset~\cite{Panayotov:2015}. 
  \item[MLS-multilingual]  From the Multi-lingual Librispeech dataset~\cite{Pratap:2020} we combine material from the Dutch (NL), German (DE) and French (FR) subsets, in equal amounts, and limited to 100 hours. 
\end{description}

We use the standard \emph{train}, \emph{dev} and \emph{eval} (a.k.a.\ \emph{test}) splits from the Librispeech sources, and generate similar splits with non-overlapping speakers for CGN, see Table~\ref{tab:datasources} for the data statistics.  
Because we do no hyper-parameter optimization, we report results just on the \emph{dev} set, except for MLS where we have joined \emph{dev} and \emph{eval} in order to have a reasonable speaker variability.

\begin{table}[ht]
  \centering
  \captionsetup{justification=centering}
  \eightpt
  \begin{tabular}{|l|c|c|c|}
    \hline
    Data source& train & dev & eval\\
    \hline
    English LS& 100\,h / 202& 5.4\,h / 40 & 5.4\,h / 40\\ 
    Dutch CGN  & 110\,h / 1573& 1\,h / 30& 7\,h / 70\\
    MLS NL/DE/FR & 100\,h / 220 & \multicolumn2{c|}{5\,h / 108 }\\
    \hline
  \end{tabular}
  \caption{Speech data source amounts (in hours)~/ number of speakers. In this research we do not use the eval partition, except for MLS.}
  \label{tab:datasources}
\end{table}

By constructing speaker and language changes artificially, there is the danger that the model will pick up session and/or data source queues confounding the task at hand.  In~\cite{Heo:2022}, the authors apply strong augmentation in training to mitigate this.  We take a different approach by making several test sets investigating this danger: 1) a test set constructed just like the train set, but with a different set of speakers; 2) a test set constructed by concatenating segments from only one speaker or language. 

\subsection{Detecting speaker changes}

We detect changes by simply looking at the decoded symbols in inference.  For Wav2vec2 experiments, we use letter decoding without language model, which can be implemented by a simple greedy search (i.e., taking the arg max of the posteriors at each frame) followed by the `CTC trick'~\cite{Graves:2006}, i.e., consecutively removing repeated symbols and blanks.  For the \emph{Speaker Label} training condition, we first sum the posteriors of all speaker classes to form a `virtual SC' posterior, which then competes with the ordinary letter classes.  In the \emph{Language Label} condition both NL, DE, and FR classes indicate a speaker change.  For Whisper experiments, we employ greedy decoding, and simply look at the generated text and interpret SC/NL/DE/FR as a change tag.  We finetune Whisper using all lowercase transcripts, except for the change tags which are all uppercase, so we expect little confusion with normal words.

\subsection{Extracting speaker embeddings}

In the \emph{Speaker Label} training condition, we want to use the input vector to the classification layer corresponding to the SC label as embedding for speaker recognition experiments.  In~\cite{Vaessen:2022} the authors found that many pooling strategies for extracting an embedding from the Wav2vec2 vector sequence work, even choosing a \emph{random} vector (instead of taking the sequence mean) can function well for speaker recognition.   We therefore hypothesize that the vector(s) associated with the virtual SC symbol can be used for an embedding related to the speaker following that SC symbol.  We found that the summed speaker posterior can peak several times in a row at inference time, even after applying the CTC trick.  Therefore, the following algorithm is used to obtain the embedding:  we consider all  embeddings from the last `space' of the previous word (or beginning of the sequence) up to the first letter of the next word.  We choose the embedding that has maximum posterior for SC to represent the speaker that follows the SC label.  

\subsection{Evaluation metrics}

For evaluating the ASR performance, we use the traditional word error rate (WER), after removing SC or language tag symbols from reference and hypothesis. 

For speaker change detection, we report performance in terms of the false positive rate \fpr\ (SC in hypothesis where no SC in reference) and false negative rate \fnr\ (no SC in hypothesis where SC in reference), after word alignment of the transcriptions including SC or language tags.  For determining the number of SCs in the reference (relevant to \fnr), we do not count the SC at the start of the utterance, since it appeared to be trivial for the transformer to predict, but still count it as a miss in the rare event it is missing in the hypothesis.  For the number potential SC moments (relevant to \fpr), we count all word transitions in the reference that have no SC. 

For speaker recognition performance, we adopt a speaker verification test protocol.  We first make a full trial list from the test data source segments, combining all possible different speaker pairs, and all same speaker pairs that originate from different sessions.  Here, a session is defined by the audio book in Librispeech.  In order to form trials at inference time, we need to find the segments corresponding to the found SC labels, but since there are false positives and negatives, this is not always possible.  We therefore work with a proxy for the speaker recognition performance, using the following approach.  At inference time, we only consider concatenated waveforms where the number of SC labels is the same in hypothesis and reference.  A selection of the full trial list is used where both segments of a trial are found in the considered waveforms.  Because we miss (potentially hard) trials this way, we report the fraction of trials from the full list.  Using the selected trials we score all embedding pairs using the cosine similarity, and report the Equal Error Rate (EER).  

For language recognition performance we measure the language error rate as the fraction of segments with the wrong language label.  A missed language label at the start of a hypothesis is counted as a language error (as well as an SCD error).  A missed language label at a speaker change that isn't a language change is not counted as a language error, and a false positive language label isn't counted as language error if the hypothesis language is equal to the reference language.  However, these last two cases do contribute to SCD errors. 

\section{Experiments}

In all experiments we finetune either pre-trained Wav2vec2 or trained Whisper models.  For Wav2vec2, we use the hyper-parameters according to the 100\,h training condition of~\cite{Baevski:2020}, specifically, a tri-stage learning rate scheduler (10\,\% linear warm-up, 40\,\% constant, 50\,\% exponential decay) with maximum learning rate of $3\times10^{-5}$ and 100k iterations.  For experiments with the XLSR-53k pre-trained model we use training tools from Fairseq, with an effective batch size of 80. For the Base-English pre-trained model we implemented our own training based on the Huggingface library~\cite{Huggingface:2019}, using an effective batch size of 8. 

For Whisper, we use Multilingual models, and implemented training with the Huggingface library.  We finetune for 4k iterations with a learning rate schedule of 400 iterations linear warmup to $10^{-5}$ followed by an exponential decay, and an effective batch size of 32.  

\subsection{Speaker Separator vs Announce}

In a first experiment we investigate the influence on the ASR performance of the addition SC target symbols.  We use the \emph{CGN-multipeaker} train set to finetune a Wav2vec2 XLSR-53~\cite{Conneau:2021} multi-lingual pretrained model.  In constructing the train set we use segments selected in the preparation~\cite{Huijbregts:2009} for NBest~2008. These are then normalized for the sound level and randomly concatenated, up to a duration of maximum 18.75\,s, making sure the concatenated segments are all from the same language variant.  By using forced-alignments of the source segments we concatenate by overlap/add aiming at a zero gap in speech. This results in a train set with a lot of SCs, on average one every 2.32\,s.  

We contrast the ``speaker separator'' condition with a training without SC labels, with SC as ``speaker announce'', and with Whisper models.  

\begin{table}[ht]
  \eightpt
  \centering
  \captionsetup{justification=centering}
  \begin{tabular}{|l|c|r|r|r|}
    \hline
    Model& cond.& WER& \fpr& \fnr\\
    \hline
    W2v2 XLSR& no SC& 14.7& 0 & 100 \\ 
    W2v2 XLSR& Sep& 14.3& 1.76& 1.28\\ 
    W2v2 XLSR& Ann& 14.5& 1.72& 1.28\\ 
    Whisper base& Sep& 26.8& 0.95& 3.83\\ 
    Whisper small& Sep& 18.0& 1.02& 0.99\\ 
    Whisper medium& Sep& 12.8& 0.84& 1.06\\ 
    \hline
  \end{tabular}
  \smallskip
  \caption{Performance for SC detection (in \%) for Dutch, measured on the dev set.  The first two rows focus on the impact to the ASR performance of introducing the SC target.  Rows 2 and~3 compare the Speaker `Separator' and `Announce' conditions, and 2, and 4--6 compare Wav2vec2 and Whisper.}
  \label{tab:exp1}
\end{table}

From the results in Table~\ref{tab:exp1} we observe that there is no big impact of the additional SC task in terms of the WER.  Word error rates seem reasonable for this dataset~\cite{N-best:2009, limsi-nbest:2009}, with Wav2vec2 XLSR (300M parameters) somewhere between Whisper small (241M) and medium (769M).  Further,  there is no significant impact of associating a SC with a speaker-homogeneous segment that follows, rather just separating speakers, on both ASR and SC performance.  From now on we will use this `Speaker Announce' approach and simply call it SC.  

\subsection{Speaker vs Channel differences}

False positive and negative rates look promising, however, there is a caveat as our dev set consists of the same kind of concatenated speech segments, and we cannot be sure the neural network doesn't simply detect channel differences.  We therefore construct a second dev set, that we label `nosc', by concatenating random segments of the \emph{same} speaker, and investigate \fpr.  This set contains the same speech content as the original `sc' dev set, but with different segment combinations.  We do the same for the train data, and merge with the original train set to form `no+sc', a training set both with and without speaker changes in a single waveform.  

\begin{table}[ht]
  \centering
  \eightpt
  \captionsetup{justification=centering}
  \begin{tabular}{|l|c|c|c|c|c|}
    \hline
    Model& train& dev& WER& \fpr& \fnr\\
    \hline
    W2v2 XLSR& sc    & sc     & 14.5 & \?1.72 & \?1.28 \\ 
             &       & nosc   & 13.5 & 12.4\? &  -   \\
             \cline{2-6}
             & no+sc & sc     & 14.4 & \?0.89 & 29.6\? \\ 
             &       & nosc   & 13.0 & \?0.04 &  -    \\
    \hline
    Whisper small& sc& sc     & 17.2 & \?1.00 & \?0.99 \\ 
                 &   & nosc   & 15.2 & \?7.72 &  -   \\
                 \cline{2-6}
                 & no+sc & sc     & 17.1 & \?0.65 & 17.2\? \\ 
                 &       & nosc   & 15.5 & \?0.02 &  -   \\
    \hline
  \end{tabular}
  \caption{Comparison of speaker change performance (in \%) for dev data with and without speaker changes, in training conditions with only speaker changes or additional single speaker concatenated files.  In the `nosc' test case, \protect\fnr\ is not defined as the number of counted SCs in the reference is zero.}
  \label{tab:exp2}
\end{table}

We can observe in Table~\ref{tab:exp2} that training with only SC can lead to high false positive rates (12.4\,\% and 7.7\,\% for Wav2vec2 and Whisper) when there are no SCs in the test data.  The value of 12.4\,\% for the dev set corresponds to one false positive per 2.38\,s, this is just about the prior of one SC per 2.32\,s in the train set!  In fact, almost all false positives occur precisely at the join in the concatenated file, so instead of a SC detector we've built a concatenation join detection system. 

This unwanted effect can be reduced by including concatenated waveforms without SC labels in the training (\fpr\ dropping to almost zero), at the cost of incurring a fair amount of false negatives (29.6\,\% for Wav2Vec2, 17.2\,\% for Whisper).  Keeping in mind that in this dataset the SC density is very high, we believe that `no+sc' training leads to more meaningful SC detection, despite the higher \fnr.  From here on, we will train with `no+sc'-style training sets.  

\subsection{Speaker Recognition with SC embeddings}

We are interested to see if we can associate an embedding with the SC labels that can function for speaker recognition.  For this, we switch to the Librispeech English data, which allows us to finetune Wav2vec2 using the smaller ``base'' models.  Using Librispeech for speaker recognition is not a very challenging task, but we consider it apt for a proof-of-concept of using SC as a speaker embedding extraction target.  We create SCs by randomly concatenating speech segments, ensuring neighboring segments are from different speakers, up to a minimum duration of 17.5\,s.  The SC density is a lot lower than before, about one SC per 24.3\,s. 


Because in this experiment we focus on the speaker recognition capabilities, we have an additional train and dev test condition `single' where we use the original speaker-homogeneous segments from the Librispeech data source.  These conditions are shown in the fist row of Table~\ref{tab:exp3}, where we can compare the WER to the original 6.1\,\%~\cite{Baevski:2020}.  This row also shows that through training separate `speaker announce targets', we can use the embedding associated with the summed-posterior SC label for speaker recognition.  A performance of approximately 11\,\% on Librispeech speakers may not seem spectacular compared to, e.g., 2.6\,\% for the more challenging VoxCeleb data~\cite{Vaessen:2022} in a very similar architecture: using the first embedding vector with cross-entropy loss.  However, we have to keep in mind that this model \emph{also} predicts letter targets for ASR, so this in a way a multi-task setup.  

\begin{table}
  \centering
  \eightpt
  \captionsetup{justification=centering}
  \begin{tabular}{|l|l|c|c|c|c|c|}
    \hline
    Train& dev& WER& \protect\fpr& \protect\fnr& EER& \% trials\\
    \hline
    single& single& 5.83& 0.00   & -& 10.6& 98.4\\
    no+sc& single & 5.63& 0.00   & -& 10.9& 99.4\\
    no+sc& sc     & 6.10& 0.24   & 11.3 & 14.2\rlap{*}& 50.7 \\
    no+sc& nosc   & 5.94& 0.69   &  - & \?4.2\rlap{*}& \?5.7 \\
    \hline 
  \end{tabular}
  \caption{Performance for the speaker recognition experiment (EER) using Wav2vec2 base pretraining (all figures in \%).  The first row is for reference, with speaker targets trained and tested using the original single-speaker utterances.  The final column is the fraction of trials from the full trial list that could be constructed using the detected SCs. The full trial list consists of $3\,651\,753$ pairs.}
  \label{tab:exp3}
\end{table}

We see in row~2 that the ASR and speaker performance are retained when we train using multiple speakers per waveform, as in the SC experiments earlier, but test with the original Librispeech dev-clean segments.  When carrying out ASR, SCD and speaker recognition all simultaneously, as shown in the third row, we incur some hit in ASR and speaker recognition performance, even though for the latter the 14.2\,\% EER cannot directly be compared to the rows above, because of the 11.3\,\% missed SCs.  Finally, we observe in the last row that using the concatenated single-speaker waveforms, a condition similar to the second row but with longer duration waveforms, the model incurs some false positives, but appears to retain speaker recognition capabilities.  The fraction of trials for this experiment is low because the second and later segments in a waveform are (correctly) not being identified with a SC, so all trials that involve these original segments are ignored.  

\subsection{Language Change detection}

In a final experiment we investigate the ability of the models to combine language and speaker segmentation with multilingual transcription.  We concatenate pairs of speech segments from MLS Dutch, German and French.  For the first segment we uniformly choose one of the three languages, and take a random sample. The second segment is selected with $p=0.5$ from one of the other languages, with equal probability.  In case the second language is the same as the first, the second segment is chosen from the same speaker with $p=0.5$.  In the transcriptions, a language label NL/DE/FR is put before the text of the corresponding segment, except in the case the second segment is from the same speaker.  In all cases we make sure the total duration doesn't exceed 30\,s.  As a consequence of the MLS segmentation, the language/speaker change rate is quite low, on average one language label per 13.8\,s, 1 SC per 27.5\,s in the train set.  Compared to earlier experiments, there now are more target letters (49) in the multilingual ASR case, due to additional (accented) letters.  

\begin{table}
  \centering
  \eightpt
  \captionsetup{justification=centering}
  \begin{tabular}{|l|c|c|c|c|}
    \hline
    Model& WER& \fpr& \fnr& LER\\
    W2v2 XLSR&      15.5  & 0.20 & 7.96 & \?1.22 \\
    Whisper small & 20.6 & 0.22 & 5.87 & 20.7\? \\
    Whisper medium& 12.7 & 0.02 & 3.52 & \?1.16 \\
    \hline
  \end{tabular}
  \caption{Language recognition error rate (LER) for several models using the MLS NL/DE/FR dataset.  All figures are in \%.}
  \label{tab:exp4}
\end{table}

The results for this experiment are in Table~\ref{tab:exp4}.  By virtue of the mixed `no+sc' training strategy, the false positive speaker detection rate can be kept reasonably low.  For reference, $\fpr=0.20\,\%$ corresponds to 1 false positive per 205 seconds, this is about 13\,\% of the SC prior.  The \fnr\ is lower than in earlier experiments, because for the test data a language change implies a speaker change, so language change detection helps SCD.  The WER for Whisper models is somewhat higher than the original models~\cite{Radford:2022}, e.g., the medium model, that we finetune here, is reported to perform at 11.7\,\%, 6.6\,\% and 8.9\,\% for MLS NL, DE and FR, respectively.  Reasons for this include our greedy decoding vs beam search with temperature fallback, the fact that our test includes language switches, and the fact that we haven't carried out any form of hyper-parameter optimization.  Also for Wav2vec2, the WER seems higher than the original research~\cite{Conneau:2021}, with 10.8\,\%, 7.0\,\% and 7.6\,\% for MLS NL, DE, and FR.  This is for the full dataset training and with LM, where we have approximately 33 hours / language and use greedy letter decoding.  Still, for the larger models we can observe low language errors while maintaining SCD and ASR capabilies. 

\section{Conclusions}

We have shown that large pretrained transformer-based ASR networks have the potential to include both speaker and language changes in their decoded output, and that the embeddings related to speaker changes can be trained to encode speaker information that is useful for speaker comparison.  Although we have seen that there are caveats in using constructed speakers changes for training such models, we believe that realistic multiparty recordings, such as AMI~\cite{Carletta:2006}, Switchboard~\cite{Godfrey:1992} or appropriate components from CGN~\cite{Oostdijk:2003}, will lead to more robust models.  This approach requires a dataset with not only speaker and/or language changes, but orthographic transcriptions as well, but for some applications these could be generated with existing ASR systems such as Whisper-large~\cite{Radford:2022}.  

\bibliographystyle{IEEEtran}
\bibliography{david-bibdesk}

\begin{thebibliography}{10}
\providecommand{\url}[1]{#1}
\csname url@samestyle\endcsname
\providecommand{\newblock}{\relax}
\providecommand{\bibinfo}[2]{#2}
\providecommand{\BIBentrySTDinterwordspacing}{\spaceskip=0pt\relax}
\providecommand{\BIBentryALTinterwordstretchfactor}{4}
\providecommand{\BIBentryALTinterwordspacing}{\spaceskip=\fontdimen2\font plus
\BIBentryALTinterwordstretchfactor\fontdimen3\font minus
  \fontdimen4\font\relax}
\providecommand{\BIBforeignlanguage}[2]{{%
\expandafter\ifx\csname l@#1\endcsname\relax
\typeout{** WARNING: IEEEtran.bst: No hyphenation pattern has been}%
\typeout{** loaded for the language `#1'. Using the pattern for}%
\typeout{** the default language instead.}%
\else
\language=\csname l@#1\endcsname
\fi
#2}}
\providecommand{\BIBdecl}{\relax}
\BIBdecl

\bibitem{Heldner:2010}
\BIBentryALTinterwordspacing
M.~Heldner and J.~Edlund, ``Pauses, gaps and overlaps in conversations,''
  \emph{Journal of Phonetics}, vol.~38, no.~4, pp. 555--568, 2010. [Online].
  Available:
  \url{https://www.sciencedirect.com/science/article/pii/S0095447010000628}
\BIBentrySTDinterwordspacing

\bibitem{Chen:1998}
S.~S. Chen and P.~S. Gopalakrishnan, ``Speaker, environment and channel change
  detection and clustering via the {B}ayesian {I}nformation {C}riterion,'' in
  \emph{Proceedings of the Darpa Broadcast News Transcription and Understanding
  Workshop}, 1998.

\bibitem{Ajmera:2004}
J.~Ajmera, I.~McCowan, and H.~Bourlard, ``Robust speaker change detection,''
  \emph{IEEE Signal Processing Lettres}, vol.~11, no.~8, pp. 649--651, 2004.

\bibitem{Yin:2017}
R.~Yin, H.~Bredin, and C.~Barras, ``{Speaker Change Detection in Broadcast TV
  Using Bidirectional Long Short-Term Memory Networks},'' in \emph{Proc.
  Interspeech 2017}, 2017, pp. 3827--3831.

\bibitem{Xia:2022}
\BIBentryALTinterwordspacing
W.~Xia, H.~Lu, Q.~Wang, A.~Tripathi, Y.~Huang, I.~L. Moreno, and H.~Sak,
  ``Turn-to-diarize: Online speaker diarization constrained by transformer
  transducer speaker turn detection,'' 2021. [Online]. Available:
  \url{https://arxiv.org/abs/2109.11641}
\BIBentrySTDinterwordspacing

\bibitem{Baevski:2020}
A.~Baevski, Y.~Zhou, A.~Mohamed, and M.~Auli, ``wav2vec 2.0: A framework for
  self-supervised learning of speech representations,'' in \emph{Advances in
  Neural Information Processing Systems}, vol.~33, 2020, pp. 12\,449--12\,460.

\bibitem{Radford:2022}
\BIBentryALTinterwordspacing
A.~Radford, J.~W. Kim, T.~Xu, G.~Brockman, C.~McLeavey, and I.~Sutskever,
  ``Robust speech recognition via large-scale weak supervision,'' 2022.
  [Online]. Available: \url{https://arxiv.org/abs/2212.04356}
\BIBentrySTDinterwordspacing

\bibitem{Pepino:2021}
L.~Pepino, P.~Riera, and L.~Ferrer, ``{Emotion Recognition from Speech Using
  wav2vec 2.0 Embeddings},'' in \emph{Proc. Interspeech 2021}, 2021, pp.
  3400--3404.

\bibitem{Tjandra:2021}
A.~Tjandra, D.~G. Choudhury, F.~Zhang, K.~Singh, A.~Baevski, A.~Sela, Y.~Saraf,
  and M.~Auli, ``Improved language identification through cross-lingual
  self-supervised learning,'' \emph{arXiv preprint arXiv:2107.04082}, 2021.

\bibitem{Vaessen:2022}
N.~Vaessen and D.~A. van Leeuwen, ``Fine-tuning wav2vec2 for speaker
  recognition,'' in \emph{IEEE International Conference on Acoustics, Speech
  and Signal Processing (ICASSP)}, 2022, pp. 7967--7971.

\bibitem{Graves:2006}
A.~Graves, S.~Fern{\'a}ndez, F.~Gomez, and J.~Schmidhuber, ``Connectionist
  temporal classification: labelling unsegmented sequence data with recurrent
  neural networks,'' in \emph{Proceedings of the 23rd international conference
  on Machine learning}, 2006, pp. 369--376.

\bibitem{Vaswani:2017}
A.~Vaswani, N.~Shazeer, N.~Parmar, J.~Uszkoreit, L.~Jones, A.~N. Gomez,
  {\L}.~Kaiser, and I.~Polosukhin, ``Attention is all you need,'' in
  \emph{Advances in neural information processing systems}, 2017, pp.
  5998--6008.

\bibitem{Snyder:2018}
D.~Snyder, D.~Garcia-Romero, G.~Sell, D.~Povey, and S.~Khudanpur, ``X-vectors:
  Robust {DNN} embeddings for speaker recognition,'' in \emph{ICASSP}, 2018.

\bibitem{Desplanques:2020}
B.~Desplanques, J.~Thienpondt, and K.~Demuynck, ``{ECAPA-TDNN: Emphasized
  Channel Attention, Propagation and Aggregation in TDNN Based Speaker
  Verification},'' in \emph{Proc. Interspeech 2020}, 2020, pp. 3830--3834.

\bibitem{Deng:2019}
J.~Deng, J.~Guo, N.~Xue, and S.~Zafeiriou, ``Arcface: Additive angular margin
  loss for deep face recognition,'' in \emph{Proceedings of the IEEE/CVF
  Conference on Computer Vision and Pattern Recognition (CVPR)}, June 2019.

\bibitem{Chan:2016}
W.~Chan, N.~Jaitly, Q.~Le, and O.~Vinyals, ``Listen, attend and spell: A neural
  network for large vocabulary conversational speech recognition,'' in
  \emph{2016 IEEE International Conference on Acoustics, Speech and Signal
  Processing (ICASSP)}, 2016, pp. 4960--4964.

\bibitem{Heo:2022}
\BIBentryALTinterwordspacing
H.-S. Heo, Y.~Kwon, B.-J. Lee, Y.~J. Kim, and J.-w. Jung, ``High-resolution
  embedding extractor for speaker diarisation,'' 2022. [Online]. Available:
  \url{https://arxiv.org/abs/2211.04060}
\BIBentrySTDinterwordspacing

\bibitem{Oostdijk:2003}
N.~H.~J. Oostdijk and D.~Broeder, ``The {S}poken {D}utch {C}orpus and its
  exploitation environment,'' in \emph{Proceedings of the 4th International
  Workshop on Linguistically Interpreted Corpora (LINC-03).}, Budapest,
  Hungary, 2003.

\bibitem{N-best:2009}
D.~A. van Leeuwen, J.~Kessens, E.~Sanders, and H.~van~den Heuvel, ``Results of
  the {N-Best} 2008 {D}utch speech recognition evaluation,'' in \emph{Proc.\
  Interspeech}.\hskip 1em plus 0.5em minus 0.4em\relax Brighton: ISCA,
  September 2009, pp. 2571--2574.

\bibitem{Panayotov:2015}
V.~Panayotov, G.~Chen, D.~Povey, and S.~Khudanpur, ``Librispeech: an {ASR}
  corpus based on public domain audio books,'' in \emph{2015 IEEE international
  conference on acoustics, speech and signal processing (ICASSP)}.\hskip 1em
  plus 0.5em minus 0.4em\relax IEEE, 2015, pp. 5206--5210.

\bibitem{Pratap:2020}
V.~Pratap, Q.~Xu, A.~Sriram, G.~Synnaeve, and R.~Collobert, ``{MLS: A
  Large-Scale Multilingual Dataset for Speech Research},'' in \emph{Proc.
  Interspeech 2020}, 2020, pp. 2757--2761.

\bibitem{Huggingface:2019}
\BIBentryALTinterwordspacing
T.~Wolf, L.~Debut, V.~Sanh, J.~Chaumond, C.~Delangue, A.~Moi, P.~Cistac,
  T.~Rault, R.~Louf, M.~Funtowicz, J.~Davison, S.~Shleifer, P.~von Platen,
  C.~Ma, Y.~Jernite, J.~Plu, C.~Xu, T.~L. Scao, S.~Gugger, M.~Drame, Q.~Lhoest,
  and A.~M. Rush, ``Huggingface's transformers: State-of-the-art natural
  language processing,'' 2019. [Online]. Available:
  \url{https://arxiv.org/abs/1910.03771}
\BIBentrySTDinterwordspacing

\bibitem{Conneau:2021}
A.~Conneau, A.~Baevski, R.~Collobert, A.~Mohamed, and M.~Auli, ``{Unsupervised
  Cross-Lingual Representation Learning for Speech Recognition},'' in
  \emph{Proc. Interspeech 2021}, 2021, pp. 2426--2430.

\bibitem{Huijbregts:2009}
M.~Huijbregts, R.~Ordelman, L.~v.~d. Werff, and F.~Jong, ``{SHoUT}, the
  {U}niversity of {T}wente submission to the {N-B}est 2008 speech recognition
  evaluation for {D}utch,'' in \emph{Proc.\ Interspeech}.\hskip 1em plus 0.5em
  minus 0.4em\relax ISCA, 2009, pp. 2575--2578.

\bibitem{limsi-nbest:2009}
J.~Despres, P.~Fousek, J.-L. Gauvain, S.~Gay, Y.~Josse, L.~Lamel, and
  A.~Messaoudi, ``Modeling {Northern} and {Southern} varieties of {Dutch} for
  {STT},'' in \emph{Proc.\ Interspeech}.\hskip 1em plus 0.5em minus 0.4em\relax
  Brighton: ISCA, 2009, pp. 96--99.

\bibitem{Carletta:2006}
J.~Carletta, S.~Ashby, S.~Bourban, M.~Flynn, M.~Guillemot, T.~Hain, J.~Kadlec,
  V.~Karaiskos, W.~Kraaij, M.~Kronenthal, G.~Lathoud, M.~Lincoln, A.~Lisowska,
  I.~McCowan, W.~Post, D.~Reidsma, and P.~Wellner, ``The {AMI} meeting corpus:
  A pre-announcement,'' in \emph{Machine Learning for Multimodal Interaction},
  S.~Renals and S.~Bengio, Eds.\hskip 1em plus 0.5em minus 0.4em\relax Berlin,
  Heidelberg: Springer Berlin Heidelberg, 2006, pp. 28--39.

\bibitem{Godfrey:1992}
J.~J. Godfrey, E.~C. Holliman, and J.~McDaniel, ``Switchboard: telephone speech
  corpus for research and development,'' in \emph{Proc.\ International
  Conference on Acoustics, Speech, and Signal Processing (ICASSP)}, 1992, pp.
  517--520.

\end{thebibliography}

\end{document}